# The independence paradox in scientific careers

**Yanmeng Xing[1,2,3], Ye Sun[4*], Tongxin Pan[2], Giacomo Livan[5,6,7†], and Yifang Ma[2‡]**

[1]Research Institute for Data Management Innovation, Nanjing University, Suzhou 215163, Jiangsu, China
[2]Department of Statistics and Data Science, Southern University of Science and Technology, Shenzhen 518055, Guangdong, China
[3]School of Information Management, Nanjing University, Nanjing 210096, Jiangsu, China
[4]School of Mathematics, Southeast University, Nanjing 210096, Jiangsu, China
[5]Dipartimento di Fisica, Universita` di Pavia, via Bassi 6, 27100, Pavia, Italy
[6]Department of Computer Science, University College London, 66-72 Gower Street, WC1A 6EA London, UK
[7]Istituto Nazionale di Fisica Nucleare, Sezione di Pavia, via Bassi 6, 27100, Pavia, Italy
[*]yesun@seu.edu.cn
[†]giacomo.livan@unipv.it
[‡]mayf@sustech.edu.cn

## ABSTRACT

Establishing an independent academic identity is a central yet insufficiently understood challenge for early-career researchers. However, limited resources and mentor-driven research agendas often constrain early efforts toward autonomy. To provide large-scale quantitative evidence on how junior researchers develop independence, we introduce a framework that traces how mentees diverge from their mentors in both research topics and collaboration networks, and how these divergences relate to long-term scientific impact. Analyzing over 500,000 mentee-mentor pairs in Chemistry, Neuroscience, and Physics across six decades, we find that high-impact scientists often initiate work in secondary areas of their mentors' expertise while adaptively establishing distinct research trajectories. This pattern is most pronounced among mentees who eventually surpass their mentors' impact. We identify an inverted U-shaped relationship between topic divergence and mentees' enduring impact, with moderate divergence yielding the highest scientific impact, revealing an independence paradox in scientific careers. This pattern holds whether topic divergence is measured by citation-network or semantic thematic distance. We further reveal that excessive direct mentor-mentee collaborations correlate with lower mentee impact, whereas expanding professional networks to include mentors' collaborators is beneficial. These findings not only offer actionable guidance for early-career researchers navigating independence but also inform institutional policies that promote mentorship structures supporting intellectual innovation and recognizing original contributions in promotion evaluations.

## Introduction

Mentorship is widely recognized as a cornerstone in shaping the academic trajectories of young scientists.[1–7]. Through close guidance, mentees often follow in the footsteps of their mentors, adopting similar research interests and professional networks. This alignment can provide a strong foundation for early career development, as mentees benefit from their mentors' expertise, networks, and established reputations[8–11]. Yet, effective mentorship extends beyond mere replication; its ultimate aim is to cultivate independent thinkers who can drive innovation and make lasting contributions to their fields[12].

A striking example of mentorship's transformative influence is the case of Giuseppe Levi, an anatomist who nurtured three Nobel laureates in Physiology and Medicine: Salvador Luria, Renato Dulbecco, and Rita Levi-Montalcini[13]. These mentees not only built upon their mentor's foundation but also diverged into distinct research fields, ultimately achieving greater recognition than Levi himself. Their success suggests that intellectual autonomy in research direction may play a critical role in career outcomes. For early-career researchers, however, achieving academic independence poses significant challenges. They must not only determine their research directions, but also consider how to collaborate with their mentors to foster a mutually beneficial relationship. On one hand, close mentorship in the early career stage offers substantial benefits: it helps mentees develop research skills[9,10,14], expand professional networks[8,11,15], increase productivity[8,16–19], improve retention[3,11,20–23], secure grants[24], and enhance career satisfaction by providing psychosocial support and work-life guidance[5,25]. These advantages foster persistence and advancement[11,26–30], but can also entrench dependence, making it difficult for mentees to establish fully independent trajectories. On the other hand, mentees' independence may be constrained by the structure of mentoring relationships, especially under growing pressures for rapid publication[31] and grant competition[32], which can create conflicts not only among peers but also between mentors and their mentees[33,34]. Large groups with unequal mentor supervision may reduce mentees' chances of survival[11,35], and restrictions on "porting" projects from mentors can limit mentees' autonomy in pursuing new research directions[36]. These tensions highlight the need for a systematic understanding how early-career researchers balance the legacy of mentorship with the pursuit of independence, and how such patterns influence their career success.

To better understand how academic independence takes shape, it is useful to disentangle its core dimensions. One central dimension concerns topic divergence—the extent to which a mentee's work departs from the research directions of their mentor. Empirical studies in STEM fields show that mentees whose research topics diverge from those of their mentors are more likely to succeed in academia[10]. Exposure to diverse expertise[37,38]–particularly from mentors with distinct disciplinary backgrounds during postdoctoral training[9,14]–can promote intellectual synthesis, foster interdisciplinary innovation, thereby enhancing career prospects[37–40]. However, divergence also entails risks. Contemporary researchers shift topics more frequently than earlier generations[41], yet frequent switching or abrupt pivot–articles that depart sharply from an author's prior agenda tend to receive fewer citations[42,43], potentially imposing longer-term costs on career development. A complementary dimension of independence is collaboration divergence, defined as the transition from mentor-dependent to self-directed networks, which similarly involves critical trade-offs. Early-career collaborations with eminent scientists[17] or award-winning mentors[10] can substantially enhance a researcher's academic impact and long-term career prospects. Yet persistent reliance on a narrow or mentor-centered network may constrain a mentee's ability to explore new directions or establish a distinctive research profile[44]. Diversifying collaborators, especially those with varied expertise, not only facilitates entry into novel domains but is also a hallmark of impactful scientists[45]. The size of collaborative groups presents its own paradox: while mentees from larger research teams often achieve superior academic performance, they also face lower survival rates[11,35], reflecting the tension between short-term resource access and long-term career sustainability in the pursuit of collaboration-based independence. Nonetheless, little is known about how researchers reshape mentor-mentee collaboration networks over time to advance their careers.

Taken together, existing research on academic independence has provided valuable insights into topic and collaboration divergence, but most studies have examined these dimensions in isolation or focused on short-term outcomes and linear effects. The fundamental trade-offs early-career researchers face—between leveraging mentor guidance and pursuing independent research—remain largely unexplored. To address these gaps, we develop a unified quantitative framework to study how mentees diverge from their mentors across both research topics and collaborations over the course of their careers, and how these dynamics jointly shape long-term scientific impact. We analyze large-scale longitudinal data in Chemistry, Neuroscience, and Physics (1960-2021), linking mentor-mentee genealogies from the Academic Family Tree[46] with publications and authorship records from OpenAlex[47]. Topic divergence is quantified by measuring thematic dissimilarity between mentor-mentee pairs, with research topics identified via community detection applied to their publication co-citation networks. Collaboration divergence is accessed through co-authorship frequency and network overlap with mentors and their collaborators. Using these metrics, we track mentees' independent trajectories and citation impact, comparing behavioral patterns between elite and general researchers. We further complement these analyses with a separate semantic-distance measure of topic divergence. Finally, we apply regression models to determine the nonlinear relationship between long-term impact and this two-dimensional independence, while controlling for field, productivity, mentor impact, and other covariates.

## Results

### A framework for measuring academic independence across topics and collaborations

We conceptualize academic independence along two complementary dimensions—topic divergence (intellectual independence, Figure 1a) and collaboration divergence (professional independence, Figure 1b)—which together provide a framework for understanding how differences from mentors in these dimensions influence mentees' scientific impact.

On the topic dimension, for each mentor–mentee pair, we first identify research topics using community detection on their papers' co-citation networks (Methods), and then classify each research area into *mentor's topics* (where the mentor has published) and *mentee's new topics* (published solely by the mentee). A mentee's topic selection can thus be summarized by the proportion of new topics among all of their topics, denoted as $R$. As shown in the middle of Figure 1a, a mentee publishes in two topics also engaged by their mentor and initiates two additional topics independently (*co-engage and innovate*), then $R = 1/2$. Values of $R$ span a continuum from *pure co-engage* ($R = 0$, mentees work only on the same topics as mentors) to *pure innovate* ($R = 1$, mentees develop entirely new agendas). To further differentiate the mentor's influence, we distinguish between mentor's *primary* and *secondary* topics: some are deemed primary if the share of the mentor's publications in that topic exceeds their median engagement across all topics, while the rest are classified as secondary. This distinction provides a finer-grained lens on mentees' choices, as primary topics reflect the mentor's core expertise whereas secondary ones represent peripheral involvement (Supplementary Fig. S11). These measures allow us to quantify the degree of topical alignment, capturing coupled mentor-mentee relationship in research topics. In addition, topic divergence can be continuously quantified as the average shortest-path distance between mentor and mentee papers within the co-citation network, see Methods for details. It should be noted that community detection is sensitive to network structure[48–50], such that networks covering topics of different breadth may yield communities at inconsistent levels of granularity. We therefore validate the robustness of this approach with a text-based semantic distance to measure topic divergence (Supplementary Note 5).

Given our operationalization of topic divergence, and drawing on prior work, we note that intellectual synthesis and topic exploration promote scientific growth[9,10,45], while large or frequent topic shifts tend to reduce citation impact[41,43] In addition, researchers working in closely related topic communities often compete for citations[51]. Together, these findings point to potential risk-reward trade-offs in intellectual independence. We therefore expect that cumulative citation impact will peak at moderate topic divergence (Figure 1a)—when mentees balance contributions to mentors' topics with their own innovations—and their independence pathway generally unfolds in stages: mentees begin in their mentors' primary topics, shift into secondary areas, and eventually establish their own research directions (Figure 1c).

On the collaboration dimension, empirical studies indicate that insufficient mentor engagement increases early attrition risk[11,35], while excessive dependence limits the formation of a distinct scholarly identity[10,44]. We thus conceptualize collaboration independence as a dynamic process (Figure 1b): mentees begin with close, direct ties to mentors, broaden into indirect links through shared collaborators, and ultimately establish self-directed networks. In practice, this can be captured by measures as the proportion of direct mentor ties or the fraction of mentor's collaborators in a mentee's ego network. Accordingly, we argue that moving from direct mentor-centered ties to more self-directed collaboration networks is conducive to sustainable career success.

Collectively, the topic and collaboration dimensions constitute an integrated framework (Fig. 1) for quantifying academic independence, which allows us to track how mentees dynamically allocate effort across mentors' primary and secondary research areas, venture into new topics, and progressively restructure their collaboration networks. Guided by this framework, we first present our network-based approach to integrating research topic detection with the measurement of scientific impact. Our empirical analyses then examine (i) whether topic divergence relates nonlinearly to long-term impact, (ii) whether topic exploration follows a staged trajectory, and (iii) whether progressively independent collaboration networks predict sustained career success. In addition, this framework enables comparisons between the independent career trajectories of all researchers and those of high-impact researchers.

### Research topic detection and scientific impact measure

In this study, we analyze genealogical data from 0.5 million mentorship records involving 80k scientists, each of whom published at least 20 articles, totaling nearly 10 million papers in Chemistry, Neuroscience and Physics between 1960 and 2021 (Methods and Supplementary Table S2). The selection of these fields is justified in Supplementary Section 1.2.1 (Supplementary Table S1 and Fig. S1). The citation profiles of mentors' and mentees' papers, along with those citing them, are extracted from the OpenAlex database.

To understand how mentees develop intellectual independence from their mentors, a direct approach is analyze their topical relationships in scientific publications. To achieve this, we construct for each mentorship pair a co-citation network (Fig. 2a). In this network, each node represents a paper authored by either the mentor, the mentee, or both, and two nodes are connected if their corresponding papers are cited together by at least one common source. This method of linking nodes is widely used to detect research topic similarity, as papers co-cited by common sources are more likely to be semantically related than randomly paired papers[52,53]. We are aware that our approach to building co-citation networks could be enhanced via a statistical validation method aimed at distinguishing statistically significant co-citations from co-citations that could effectively be considered random events[54,55]. At any rate, we validate our co-citation networks to be made of communities that can be meaningfully associated with different research topics (Supplementary Fig. S10-S12).

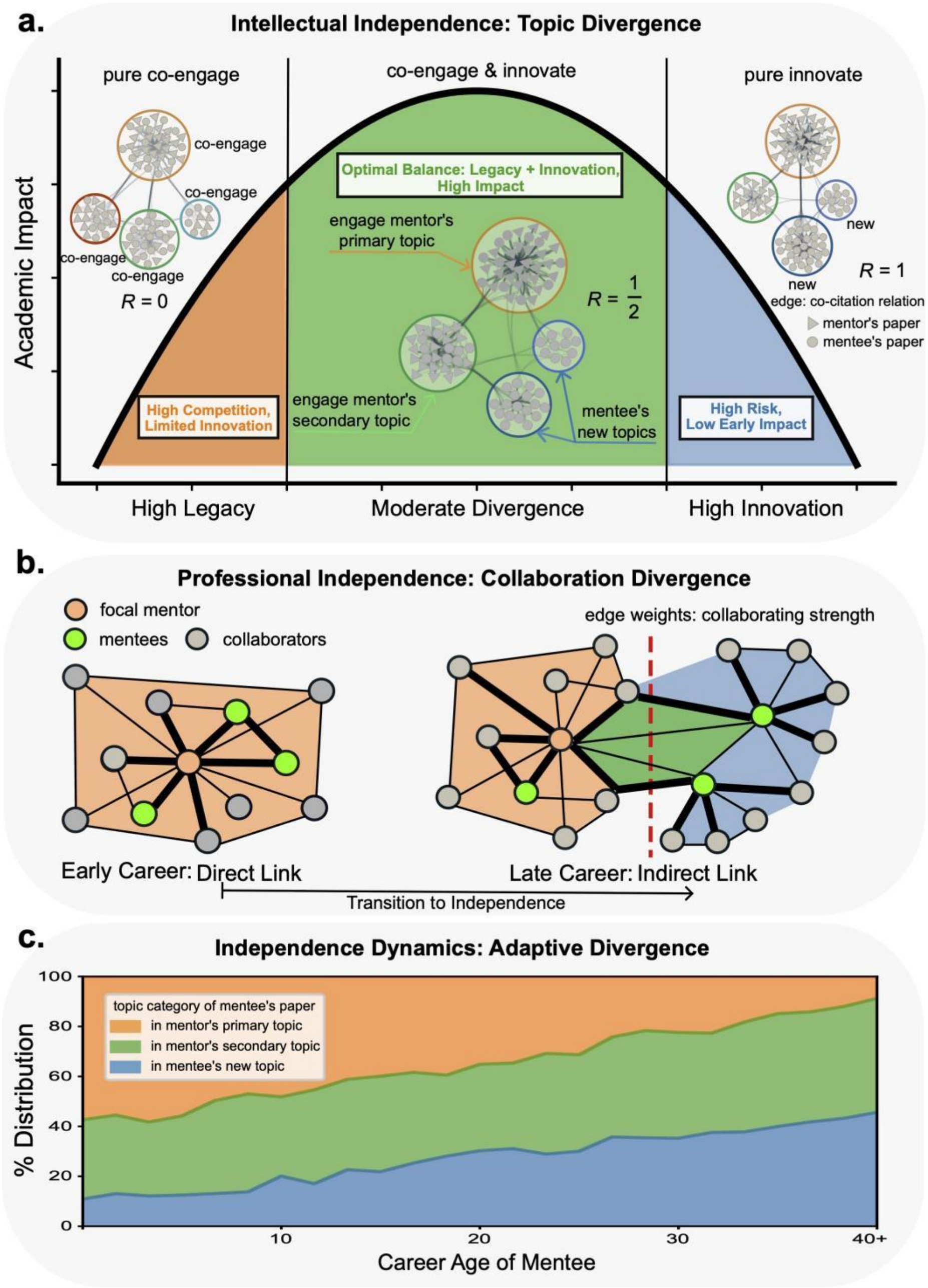

**Figure 1.** **Conceptualizing academic independence through the evolution of topic and collaboration divergence. a.** Intellectual independence: topic divergence. Research topics are identified using a co-citation network constructed from the published papers of both the mentee and mentor, with each community representing a distinct topic. The graph ranges from closely following the mentor's primary (high legacy, limited innovation) or secondary topics to entirely exploring new topics (higher risk, lower early impact). **b.** Professional independence: collaboration divergence. The network visualizes the mentees transitioning from early to late career stages through changes in collaboration patterns. In the early career phase, mentees collaborate mainly with their mentors and the mentors' network (orange shade). As they progress to the late career stage, they shift to indirect collaborations (green shade) and form their own independent networks (blue shade). **c.** Independence dynamics: adaptive divergence. This stacked area chart shows how the mentee's topic categories evolve over their career. Mentees initially focus on the mentor's primary topic (orange), gradually shifting toward secondary topics (green) and new areas (blue).

The topic communities in each co-citation network are identified with the fast unfolding algorithm[56], a popular approach for detecting community structures in large networks[57,58] (Methods). Typically, a co-citation network consists of several large topic communities, along with some smaller clusters and isolated nodes. To ensure the identified research topics are meaningful[41,50,59,60], we include only topic communities with at least 10 nodes in this study. Following this procedure, the co-citation network also retains the majority of papers with citations from both the mentor and mentee (Supplementary Fig. S2). Fig. 2a illustrates a co-citation network of a typical mentor-mentee pair, featuring 6 distinct topic communities, with each node coloured according to its identified topic community. Nodes are shaped as circles, triangles or triangles inside circles, representing papers authored by the mentee, mentor, or both, respectively. Fig. 2b illustrates the evolution of the research focus for the mentor-mentee pair, highlighting their respective publications on different topics throughout their careers. The downside of this approach is that cluster results are inevitably shaped by the underlying citation network structure[48–50]. When a co-citation network spans multiple subfields, fields, or even domains, community detection may be driven more by structural artifacts than by true topical similarity. Consequently, the detected communities may correspond to topics at different levels of granularity—that is, some reflecting narrow subfields while others capture much broader fields—which limits comparability across networks of different scope. To check this, we compared the detected communities with the OpenAlex topic classification and confirmed that most communities indeed align with research topics at a consistent level of granularity (Supplementary Fig. S3, S10-S12).

To measure the research impact of a mentor-mentee pair, we adapt and apply a collective impact allocation algorithm[33,61] to capture their respective contributions to the papers within each topic community. Fig. 2c presents an example of how we calculate the impact of a focal mentee with papers belonging to different detected topic communities. We first identify all the papers published by the mentee within a community $j$, forming a set $E_j \equiv \{e_{j,1}, e_{j,2}, \dots, e_{j,n_j}\}$, and all the papers published by the mentor, forming a set $R_j \equiv \{r_{j,1}, r_{j,2}, \dots, r_{j,m_j}\}$. Here, $n_j$ and $m_j$ respectively represent the number of papers authored by the mentee and mentor in the community $j$. The complete set of all papers published by the mentee and mentor in community $j$ is $M_j = E_j \cup R_j$. We then identify all co-citing papers $P_j$, which comprises the complete set of papers co-citing at least two papers together from the set $M_j$. Based on this framework, the total impact $C^e$ accrued by a *mentee* from all their papers $E_j$
within community $j$ is defined as:

$$C_j^e = \sum_{i=1}^{n_j} \frac{w_{e_{j,i}}}{s_{e_{j,i}}} \tag{1}$$

where the impact $w_{e_{j,i}}$ is the number of citations that paper $e_{j,i}$ has received from the co-citing papers $P_j$. The term $s_{e_{j,i}}$ represents the number of authors on the mentee's paper $e_{j,i}$, which serves to normalize the impact $w_{e_{j,i}}$ and fairly distribute the impact among co-authors[62]. Similarly, the total impact $C^r$ attained by a *mentor* from all their papers $R_j$ within community $j$
can be expressed as:

$$C_j^r = \sum_{i=1}^{m_j} \frac{w_{r_{j,i}}}{s_{r_{j,i}}} \tag{2}$$

where $w_{r_{j,i}}$ signifies the number of citations received by paper $r_{j,i}$ of the mentor from the co-citing papers $P_j$, while $s_{r_{j,i}}$ represents the number of authors on paper $e_{j,i}$. In sum, this framework integrates topic detection with impact assessment to quantify the separate contributions of mentors and mentees within topic communities, providing a reproducible and generalizable measure of intellectual independence and its influence.

### Elite mentees strike a balance between research on their mentors' topics and exploring new ones.

In the framework of analyzing academic independence in terms of topic entanglement, we explore two fundamental questions: What are the characteristic patterns of topic interaction between mentors and mentees? And which pattern better supports the mentee's career success? Additionally, we investigate whether elite mentees follow distinct interaction patterns compared to general mentee population (all mentees). The elite mentees are defined as those whose total number of citations of papers over their careers ranks within the top 20% of all mentees. We also verified the robustness of the results reported below using alternative thresholds (5%, 10%, 15%, Supplementary Section S4.2). Here we measure citation counts within five years after publication to mitigate cumulative citation bias[63] and ensure that elite mentees are comparably represented across eras (Supplementary Fig. S4). Notably, both general and elite mentees are required to have published their first papers before 1990, guaranteeing that their academic careers can span over 30 years. This criterion enables us to examine the landscape of mentor-mentee topic engagement throughout their entire career. Insets in Fig. 3a-c illustrate the distribution of mentees across topic choice patterns. We can observe that a larger fraction of all mentees adheres to the *"pure co-engage"* mode compared to elite mentees. The majority of elite mentees adopt the *"co-engage and innovate"* mode, with the orange bar significantly exceeding the purple bar, indicating that elite mentees are more likely to contribute to both shared and new topics. Although both all mentees and elite mentees have the smallest fractions in the *"pure innovate"* category, elite mentees are slightly less represented here than all mentees. Additionally, we examine the complementary cumulative distribution function (CCDF) of $R$, where the y-axis represents the probability that the proportion of a mentee's engagement in new topics exceeds $R$ (Fig. 3a-c). The results indicate crossover points between 0.7 and 0.8 in the distributions, at which the probabilities of elite and all mentees are equal. Before this point, elite mentees are more likely to engage in new topics with a ratio greater than $R$. This further reveals that the majority of elite mentees prefer a balanced approach, combining their mentors' topics with new and independent ones *"co-engage and innovate"*, Supplementary Fig. S5). These patterns are consistent across the three research fields of Chemistry, Physics, and Neuroscience.

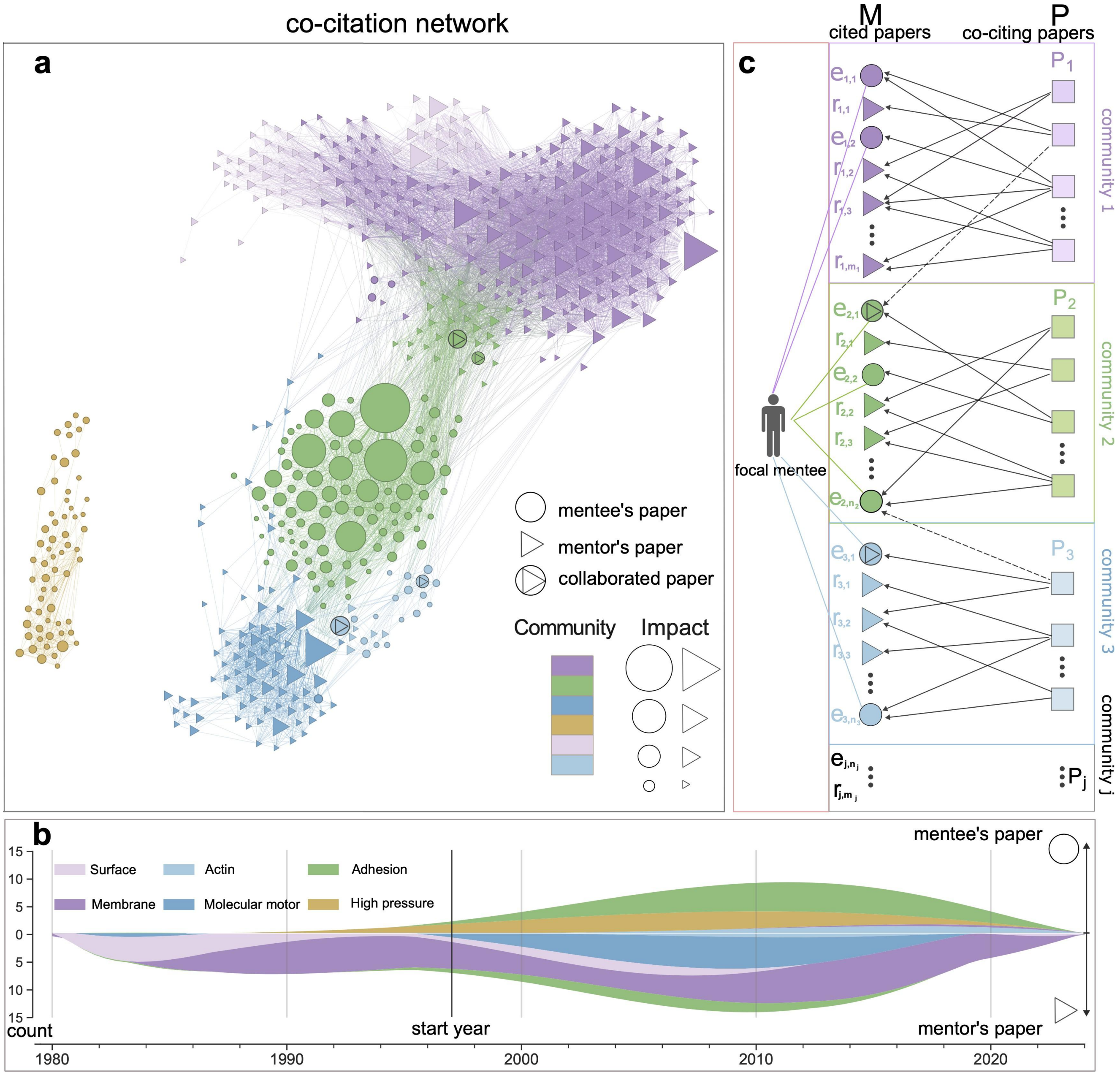

**Figure 2. Illustration of a mentor-mentee co-citation network and topic-specific impact measurement. a.** The co-citation network contains all the papers published by a mentor-mentee pair. Each node represents a paper: circles for mentee papers, triangles for mentor papers, and overlapping shapes for mentor-mentee collaborated papers. Links connect nodes if the corresponding papers are co-cited by at least one common paper. Node colors denote the different communities identified by the fast unfolding algorithm[56], and node sizes correspond to their citations. **b.** The time series tracks the network's evolution, depicting the growth of papers by the focal mentee (upper half) and the pair-wise mentor (lower half) throughout their careers. Each shaded area's color matches a topic community in the co-citation network, with the height at each year point indicating the number of papers published on that topic. The vertical black line represents the year the mentor started supervising the mentee. **c.** The heuristic algorithm quantifies the impact of a focal mentee by analyzing co-citations within each topic community. The mentors' papers ($r_{j,i}$) and mentees' papers ($e_{j,i}$) are organized by their topic communities in the left column, color-coded to align with the communities $j$ identified in panel a. Squares in the right column signify papers that co-cite any of the mentor-mentee papers, with colors matching the communities of mentor-mentee papers. Solid lines connect papers with the same colors, showing co-citations within communities, while dashed lines indicate cross-community co-citations

To gain deeper insights into how mentees engage with their mentors' research topics, we analyze the distribution of their research efforts across mentors' primary and secondary topics, as well as in their own new topics, as illustrated in Fig. 3d-f. Our analysis reveals that compared to the general mentee population, elite mentees in all three fields exhibit a consistently higher level of diversification under each topic type, particularly in secondary and new topics. This indicates that elite mentees tend to diversify their research efforts more extensively. In Chemistry and Physics (Fig. 3d and 3f), elite mentees show a particularly strong tendency to engage in new topics, suggesting a higher propensity for innovation and exploration beyond their mentors' established areas. Additionally, we find that among mentees who co-engage in their mentors' research, both elite and general mentees contribute less to mentors' primary topics than with secondary ones.

Given the above findings on mentees' engagement in three topic types, we are interested in determining which topic selection patterns yield the highest scientific impact for mentees. To this end, we calculate the average ratio of scientific impact attributed to the mentees' publications within each topic type. As shown in Fig. 3g-i, both elite and all mentees, regardless of whether they choose a *"co-engage and innovate"* or *"pure co-engage"* mode, achieve the lowest impact ratio when engaging in their mentors' primary topics, and this pattern is consistent across all three fields. Furthermore, elite researchers who adopt the *"co-engage and innovate"* mode exhibit an interesting balance: they achieve comparable impact across both new topics and their mentor's secondary topics. This suggests that maintaining such a balance is a key factor in achieving high scientific impact. In contrast, all mentees shows a higher impact ratio in new topics compared to their mentors' secondary topics. We further analyze the probability that mentees (both elite and all) publish their single most impactful paper under each topic type (Supplementary Fig. S7), arriving at similar conclusions. We also confirm that all these findings remain robust when topical breadth of co-citation network is carefully controlled (Supplementary Fig. S13-S15).

### Mentees transition from mentors' topics to independent research and eventually surpass them

We next extend our analysis to examine how mentees accumulate scientific impact within each topic type and transition across topics throughout their careers. Fig. 4a-c shows the average cumulative scientific impact of all mentees (solid black line) and their mentors (dashed black line) as a function of their career ages. The results indicate that mentors consistently exhibit higher cumulative impacts than their mentees at the same career stages in Neuroscience and Physics, while the mentees and their mentors in Chemistry show indistinguishable cumulative impacts at equivalent career ages. We then trace how mentees accumulate scientific impact within each topic type over careers, represented by orange lines for primary topics, green for secondary, and grey for new topics. We can see that, the general mentee population accumulates a similar amount of impact in their mentors' primary and secondary topics during the early-career stage (0-10 career years). As time progresses, the impact from secondary topics significantly overtakes that from primary topics after the mid-career stage (10-20 career years). This trend continues, with the gap between primary and secondary topics widening until the end of their careers. The mentee's cumulative impact on new topics (grey lines) gradually grows throughout their career. While it starts at the bottom during the early stage, it eventually surpasses the impact of primary topics in the later stage (post-20 career years). On the contrary, the corresponding mentors tend to accrue the most citation impact in their primary research topics, rather than in secondary topics (Supplementary Fig. S8). The inset boxplots in these panels offer a detailed decade-by-decade breakdown of publication distributions by topic type across all three fields.

Since some mentees eventually surpass their mentors in total impact (Fig. 4a), we now shift our focus to analyze whether topic engagement behavior differs between the general mentee population and those exceptional mentees who both rank in the top 20% by citation impact and ultimately outperform their mentors (outperforming mentees, Fig. 4d-f). Compared to all mentees, outperforming mentees begin their careers with a strong focus on secondary topics. We can see the green lines rise sharply, indicating these mentees' tendency to quickly establish significant contributions to their mentors' secondary topics. This early divergence from their mentors' primary research is evident across Chemistry, Neuroscience, and Physics. During the early stages of their careers, these outperforming mentees initially accumulate nearly equal citations from both their mentors' primary topics and their own new topics. However, they later gain more impact from new topics than from their mentors' primary topics and achieve this shift earlier in their careers than the general researcher population. The successful career trajectories of these outperforming researchers highlight the potential rewards of a balanced and innovative topic selection strategy, providing a possible roadmap for aspiring academic success in science.

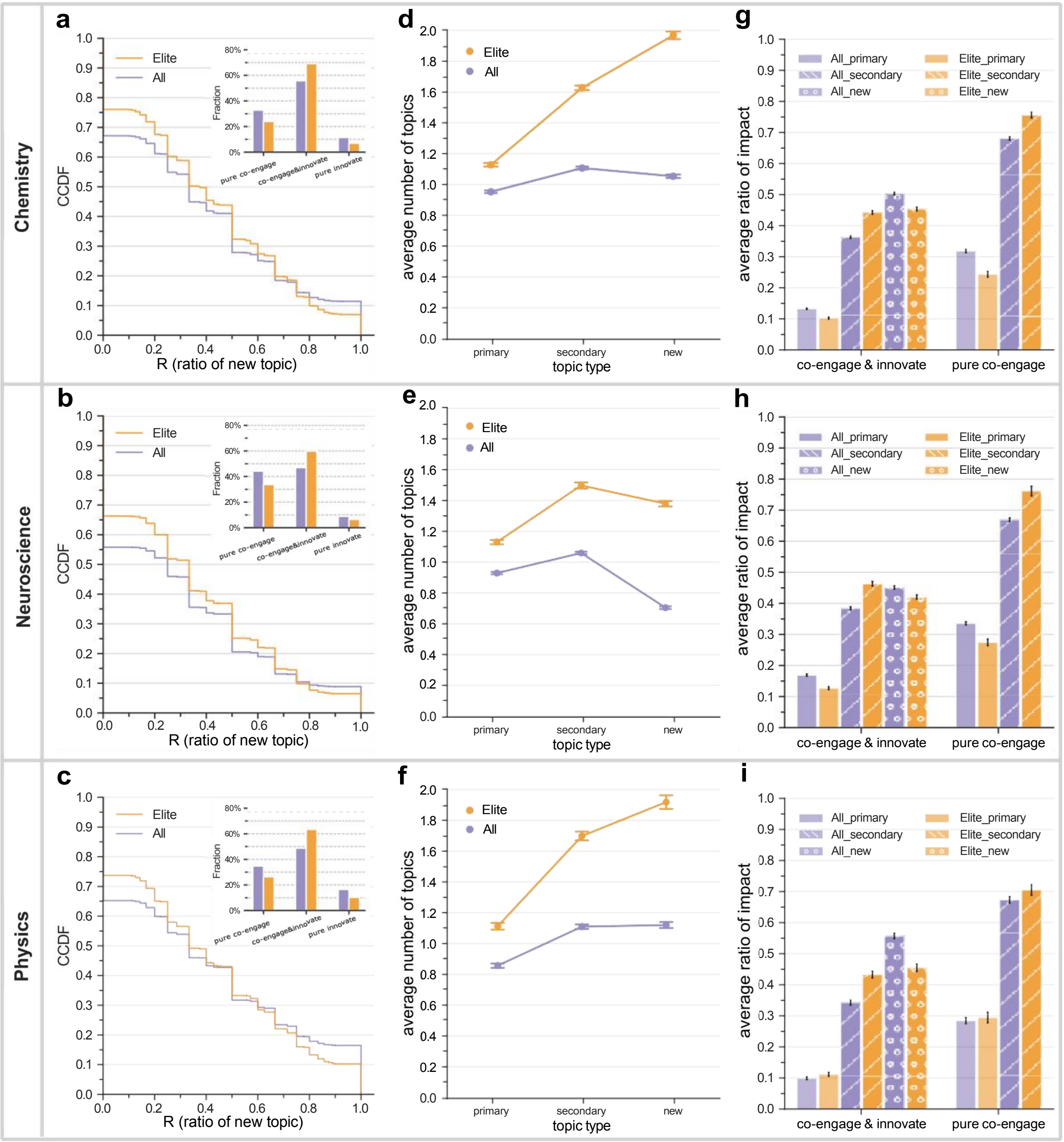

**Figure 3. Interaction patterns of research topics between mentor and mentee. a-c.** The complementary cumulative distribution function (CCDF) of the ratio of new topics ($R$) for mentees. The purple stepped line represents the CCDF ($R$) for all mentees in our dataset, while the yellow stepped line indicates the CCDF ($R$) for elite mentees, defined as those in the top 20% by cumulative citations. Inset: the fraction of mentor-mentee topic interaction patterns in panel **a** of Fig. 1 for all (purple bars) and elite mentees (yellow bars). **d-f.** The average number of topics of different types (primary, secondary and new) pursued by all mentees and elite mentees over their careers (see the distribution of the number of mentees' topics across different types in Supplementary Fig. S6). Data are presented as mean ± s.e.m. **g-i.** The average ratio of impact across different topic types for all mentees and elite mentees over their careers. This measure is computed and compared separately for all mentees and elite mentees with two topic selection patterns, namely, "co-engage & innovate" and "pure co-engage". Data are presented as mean ± s.e.m.

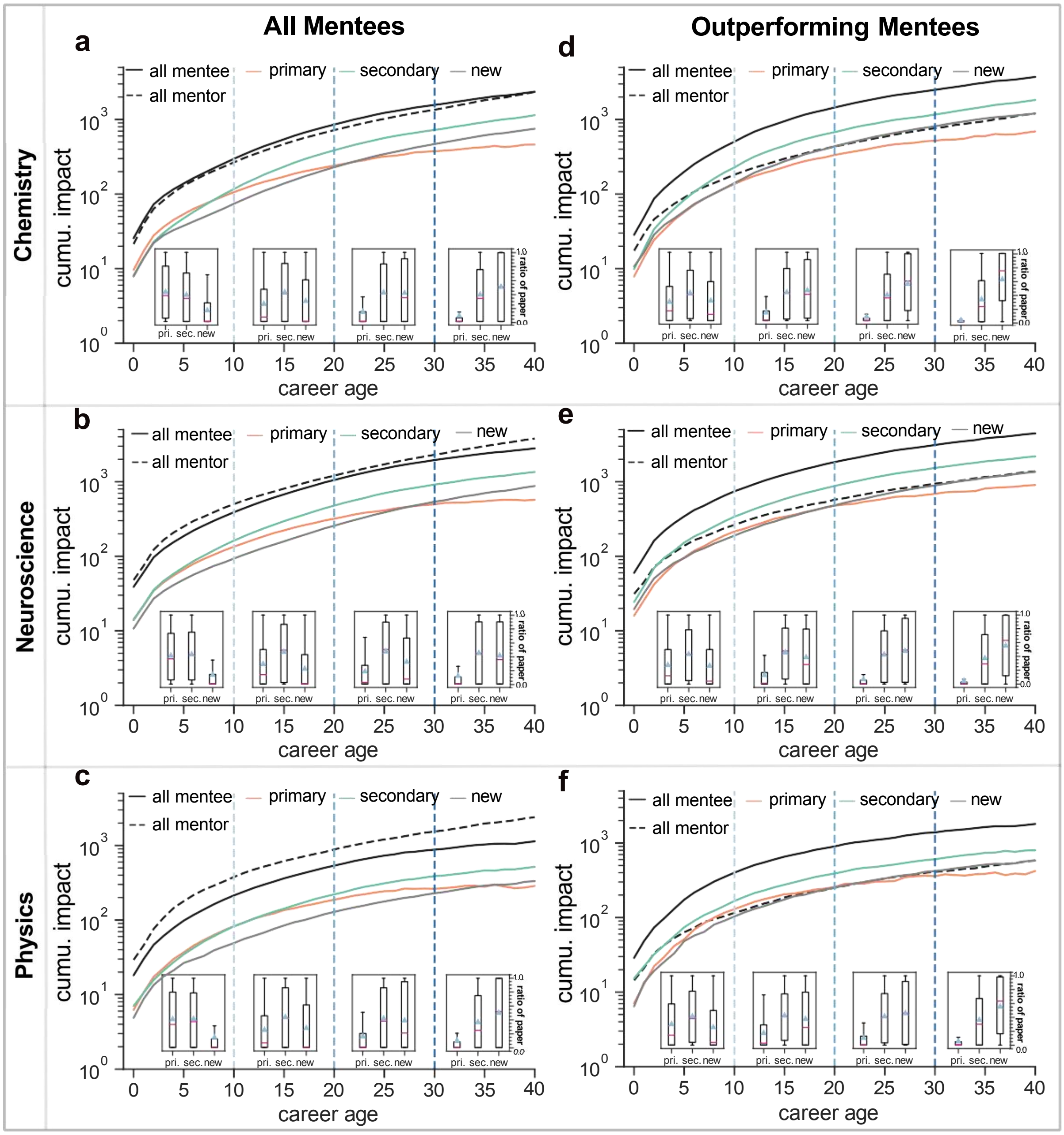

**Figure 4. Evolution of cumulative impact across different topic types for all mentees (a-c) and outperforming mentees (d-f).** The solid black line represents the average cumulative impact earned from mentees' papers, while the dashed black line shows the average cumulative impact of their corresponding mentors' papers over the career years since their first publication, respectively. The colored lines–orange for primary topics, green for secondary topics, and grey for new topics–track the average cumulative impact evolution for mentees' papers that are dedicated exclusively to these specific topic areas. The insets display boxplots for the ratio of the number of mentees' papers by topic type for each decade, with blue triangles denoting the mean and purple lines indicating the median number of papers.

## Mentees often outperform mentors in impact within their moderately similar research topics.

It remains unclear which topic selection patterns particularly enable mentees to outperform their mentors. To address this, we first categorize the mentees into four quadrants based on whether their citation impact in primary or secondary research topics exceeded that of their mentors (Fig. 5a-c). All mentees (depicted in purple points) shows a significant concentration in the third quadrant, with 60% for Chemistry, 61% for Neuroscience, 56% for Physics, indicating that the majority of mentees do not outperform their mentors in either primary or secondary topics over their careers. Followed by the proportion of mentees in the second quadrant, approximately 37% of mentees achieve more citation impact in secondary topics but less impact in primary topics than their mentors in Chemistry. A smaller proportion of mentees, specifically 2% for Chemistry and Neuroscience and 5% for Physics, are found in the first quadrant, where they outperformed their mentors in both primary and secondary topics. Lastly, only 1% of mentees are located in the fourth quadrant, with more impact in primary topics and less in secondary topics, which highlights the difficulty of exceeding mentor's impact in their established fields. In contrast, outperforming mentees (represented by yellow points) are more likely to be located in the second quadrant, exceeding their mentors' impact in secondary topics while lagging in primary topics. A smaller fraction of mentees, however, surpass their mentors in both primary and secondary topics, as shown by the yellow points in the first quadrant, representing only 14%, 20%, and 30% of mentees in Chemistry, Neuroscience, and Physics, respectively. We further observe that an even smaller subset of outperforming mentees fails to exceed their mentors' impact in either primary or secondary topics in the third quadrant. This observation motivates us to further examine their performance in new research topics.

We therefore represent each mentee on a triangular plot, visualizing their contributions across primary, secondary, and new topics (Fig. 5d-f). A mentee is positioned at the top corner if all their citations originate from primary topics. The bottom-left corner represents exclusive impact in secondary topics, while the bottom-right corner indicates impact received only from new topics. For mentees receiving citations from multiple topic types, their position within the triangle corresponds to the relative proportion of citations from each type. For instance, a mentee would be in the center if they received an equal number of citations from all three topic types. We find that the outperforming mentees, marked by yellow circles, tend to spread along the secondary-new and primary-secondary edges, rather than in the middle of the triangle. Also, very few elite mentees are positioned along the primary-new edge of the triangle. In contrast, for the general mentee population (purple circles), some mentees cluster near the corners, while many others are more evenly distributed throughout the triangle. Overall, this pattern further indicates that outperforming mentees tend to distinguish themselves through concentrated advances in mentor's secondary and their new topics, whereas the general mentee population engages relatively evenly with primary, secondary, and new topics.

The observed patterns of mentor-mentee topic engagement point to the central question: how does the degree of topic divergence between a mentee and their mentor shape the mentee's potential to achieve high impact or surpass the mentor's influence? To this end, we quantify the topic distance between a pair of mentee and mentor by calculating the average length ($L$, formula 4) of the shortest path linking their papers within the co-citation networks. A greater topic distance indicates higher dissimilarity between the mentor's and mentee's research topics. Fig. 5g-i shows the average cumulative citations of mentees as a function of the topic distance between the mentee and their mentor. Across Chemistry, Neuroscience, and Physics, we observe consistent inverted-U relationships for the all mentees, as supported by statistical tests for the fitted curve (Supplementary Table S3-S5). This result confirms the hypothesis that there is an optimal balance between leveraging a mentor's expertise and forging an independent research path, a balance that enables mentees to maximize their long-term impact. However, it does not escape us that our network-based method to quantify topic divergence may be inevitably biased by citation structure[48–50] (Supplementary Note 5). To address this limitation, we verified that the inverted-U relationship persists across three disciplines using an independent framework based on semantic topical distances between mentors' and mentees' papers (Methods, Supplementary Fig. S24), which does not rely on co-citation networks.

## Effective use of mentors' collaboration networks supports mentees' development.

To investigate the role of mentor-mentee collaboration divergence in shaping long-term academic success, we conducted non-linear regression analyses relating mentees' cumulative impact to both topic divergence and collaboration patterns, while sequentially controlling for mentee-specific and mentor-specific factors that also influence career outcomes[8,9,20,41,43,64] (Supplementary Section S3.1). The analysis was restricted to mentees with careers of at least 30 years, providing a comprehensive perspective on the evolution of research and collaboration dynamics. Regression results confirm the inverted-U relationship for topic divergence across the three fields: the linear effect of average network distance (*ave_distance*) is positive, while the quadratic term (*ave_distance*$^2$) is negative (Fig. 6; Supplementary Tables S6-S8, Model 10). Collaboration patterns reveal a nuanced temporal effect. Continued reliance on direct mentor co-authorship beyond the first five years (*colla_work_count_later*) negatively affects long-term cumulative impact in Chemistry and Neuroscience (Fig. 6a-b). In Physics, where mentees tend to maintain higher direct mentor collaboration early in their careers, over-reliance in the first five years (*colla_work_count_first_5y*) reduces cumulative impact, reflecting discipline-specific dynamics (Fig. 6c, Supplementary Fig. S9). By contrast, the number of shared collaborators (*common_collaborators_count*) between mentors and mentees consistently exhibits a positive association with cumulative impact across all fields.

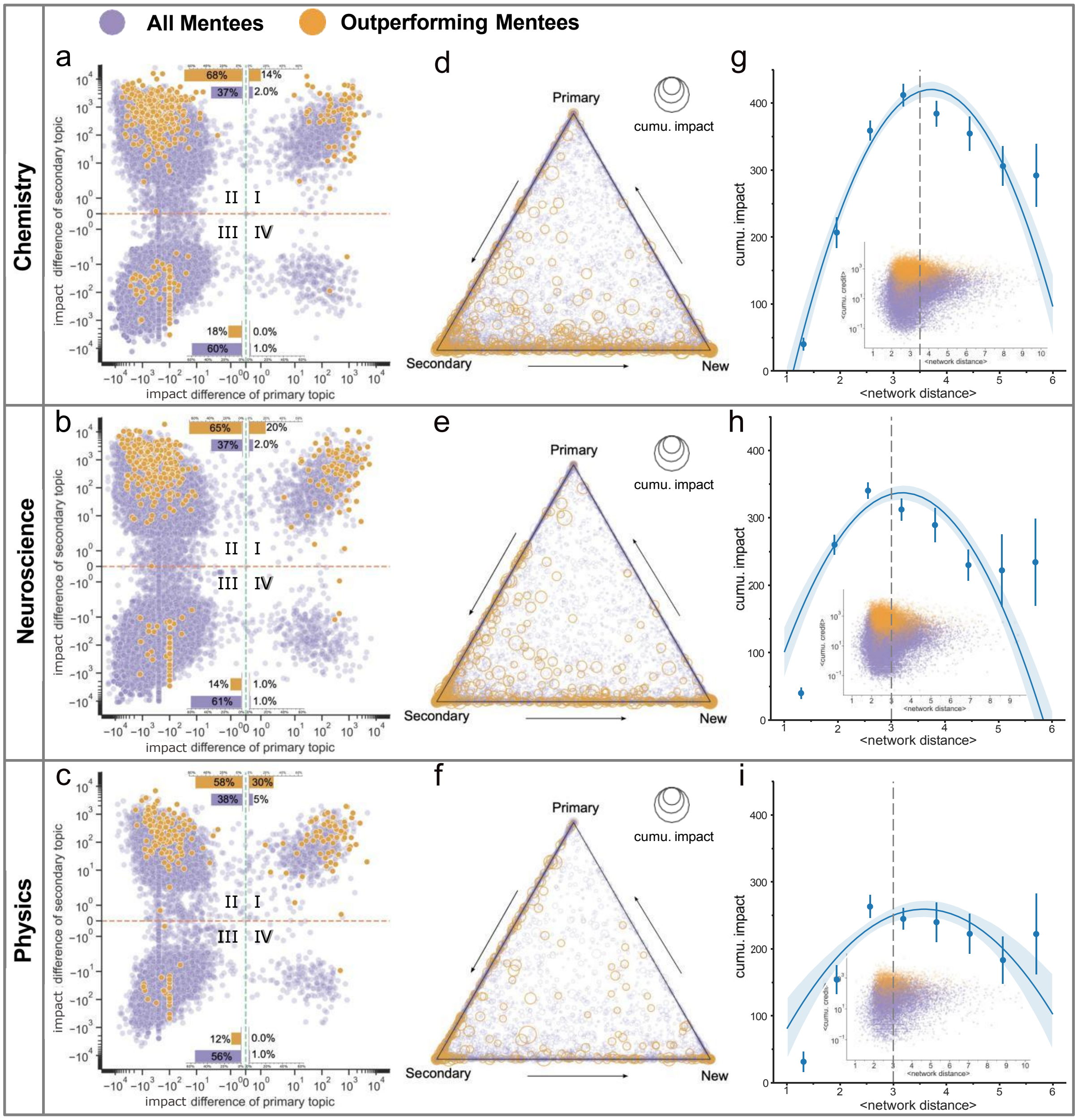

**Figure 5. Relationship between the likelihood of mentees outperforming their mentors and their research topic divergence. a-c.** Impact difference between mentees and their paired mentors in primary and secondary topics. The horizontal axis shows the impact difference for primary topics and the vertical axis for secondary topics, both defined as the mentee's impact minus the mentor's. Yellow points indicate outperforming mentees, and blue points depict all mentees. The insets, one per quadrant, report the percentage of points that fall into each quadrant for the respective group of mentees distinguished by the color of the nodes. **d-f.** The positions of mentees in the primary-secondary-new topic triangle are determined by the relative share of citations they receive from each topic type. The circle size denotes the total number of citations received by mentees, and the color denotes the group of mentees: yellow for outperforming mentees; and purple for all mentees. **g-i.** The inverted-U relationship between the topic divergence of mentees and mentors and mentees' cumulative impact. Here, the topic divergence is measured by the average length of the shortest path (<network distance>) between the mentor and mentee nodes in the co-citation network, as calculated by the formula 4. Each blue point represents the average cumulative impact corresponding to a given average path length (x-axis), with error bars showing the standard error. The blue line shows the quadratic fitting curve, and the shaded area denotes the 95% confidence interval.

Together, these results empirically support that sustainable academic success involves leveraging mentors' professional networks while progressively transitioning from direct mentor-centered ties to more self-directed collaboration networks (Fig. 1b). Moreover, several covariates are significantly associated with career outcomes: longer mentee careers (*career_len_mte*), higher productivity (*paper_num_mte*, Supplementary Table S11), and greater mentor citation impact (*total_cits_mto*), all positively correlate with cumulative impact. This suggests that the mentee's individual efforts and the mentor's reputation significantly contribute to the mentee's career development[17,65]. By contrast, mentees' long-term impact tends to decrease when mentors engage in a larger number of research topics (*topic_num_mto*), indicating that excessively broad interests may dilute the support and guidance available to early-career researchers[11,35]. Importantly, we applied the same regression framework using semantic distance instead of network distance as the independent variable. The consistent results demonstrates that our conclusions are robust and not artifacts of community detection or other structural features of co-citation networks (Supplementary Table S12-S16.)

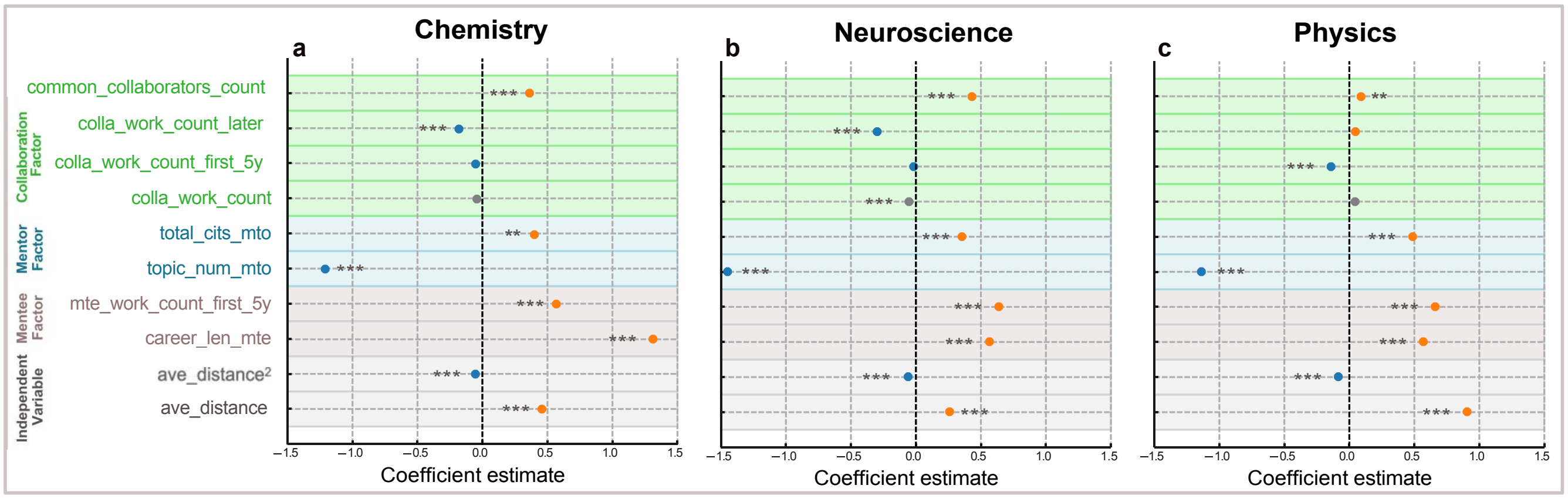

**Figure 6. Regression results of the association between the average network distance and the mentees' cumulative impact in Chemistry (a), Neuroscience (b) and Physics (c).** Results from discipline-specific OLS regressions (Ordinary Least Squares, $R^2 = 0.202, 0.203, 0.162$ for Chemistry, Neuroscience, and Physics, respectively, Model 10 in Supplementary Table S6-S8), whose dependent variable is the mentees' cumulative impact. Note that the coefficients (grey nodes) for *colla_work_count* are derived from Models 7 presented in Supplementary Table S6-S8. The statistical significance of each regression coefficient is indicated to the left of its corresponding value (* $p < 0.05$; ** $p < 0.01$; *** $p < 0.001$).

## Discussion

This study systematically compares the career trajectories of all researchers with those of elite researchers (top 20% by citation impact), focusing on divergence in research topics and collaboration networks relative to their mentors. The results reveal clear patterns in how academic independence shapes long-term scientific impact. Analysis of topic divergence shows that both general and elite mentees are more likely to succeed by working in their mentors' secondary areas or by pioneering new directions than by remaining in mentors' primary fields. Elite mentees often follow a characteristic path: they first contribute to secondary topics shared with their mentors, then branch into novel domains, and in doing so surpass their mentors' impact. This suggests a dynamic of adaptive intellectual independence, where mentees repurpose mentors' knowledge legacy to forge new research directions. Regression analyses confirm a nonlinear inverted-U relationship between topic divergence and mentees' impact, even after accounting for mentees' productivity and citations, mentors' impact, and other confounders (Supplementary Section S3.2 and S4.3). This pattern holds whether topic divergence is measured by citation-network distance or semantic distance. Together, these results suggest that moderate divergence is advantageous, but excessive departure from a mentors' topics can undermine long-term impact—a dynamic that reflects an independence paradox in scientific careers. Collaboration patterns echo this dynamic: heavy reliance on mentors' direct ties impedes independence, whereas selectively expanding professional networks—especially by engaging mentors' former collaborators—tends to boost long-term performance. Overall, these findings shed light on the complex interplay between mentorship guidance and academic independence, highlighting the pathways through which mentees evolve into independent scholars capable of innovating and advancing disciplinary boundaries.

While providing valuable insights into academic mentorship, these results have several limitations. The data sources, the Academic Family Tree and OpenAlex, are extensive but do not represent all mentor-mentee relationships across disciplines and cultures, potentially introducing selection bias[46]. Nevertheless, the AFT dataset remains the most comprehensive mentorship resource currently available in scientometric research, and prior reviews have not identified systematic biases likely to affect our results[9,10,46,66]. This study's citation-network methodology tends to overlook works with fewer citations that may nonetheless hold long-term significance[67,68], thereby biasing the analysis toward more popular topics[69]. At the same time, our measure of career achievement relies solely on citation-based indicators. Although we verified robustness across thresholds for defining elite researchers (Supplementary Section S4.2), such metrics inevitably miss other dimensions of scientific ability[70], including research design, independent thinking, supervision, collaboration, and scholarly communication. In addition, the inverted-U effect of topic divergence is necessarily validated on scholars with sufficiently long-lived careers, and we cannot quantify mentees who may have attempted early independence but subsequently left academia. The observed tendency—particularly among outperforming mentees—to move away from mentors' primary topics may reflect the broader rise of topic switching across scientific careers[41]. The costs associated with excessive topic divergence—the declining portion of the inverted-U—likely arise from the "pivot penalty" in individual papers of large topic shifts[43], compounded by broader constraints such as reduced access to mentors' networks and resources. These pivots may reduce early paper-level citations, yet their long-term effects on career trajectories remain uncertain. Finally, external factors such as institutional prestige[34], socioeconomic background[71], collaborator expertise[72], geographical location[73], and gender disparities[74,75] may interact with academic independence to influence a mentee's career success. These aspects were not comprehensively analyzed in this study. Overall, these limitations underscore that career dynamics are highly complex and cannot be fully captured by any single factor.

Our findings have profound implications for the academic ecosystem at multiple levels. They first suggest a possible roadmap for individuals towards above-average academic success: early-career researchers may benefit from starting in their mentors' secondary fields and gradually establishing independence, either within these areas or by branching into new directions. It should be noted that the path entails both greater risks and potential rewards, as frequent topic switching or exploring novel areas can lower early impact[41–43]. This is reminiscent of the trajectory observed for many researchers attempting to launch an interdisciplinary career[76]. Therefore, developing substantial independence from one's mentors is rewarding in the long run, provided that early setbacks are navigated. A successful career requires mentees to balance leveraging mentors' intellectual legacies and social capital with cultivating independence, a "sweet spot" at the core of the independence paradox. In this context, evaluation and reward systems should align incentives with long-term scientific value, since novel exploration often takes time to yield substantial outcomes[77,78]. Recognizing early signs of independent innovation among junior researchers is particularly important. Such contributions often foreshadow future success.

From a broader perspective, scientific progress advances through successive generations of researchers, who may deepen established knowledge, push disciplinary boundaries, or do both[79–82]. Our findings underscore the need to foster inclusive mentorship structures and dynamic research agendas. Such environments enable individuals to balance building on legacy with pursuing innovation throughout their careers. Supporting early-career researchers in exploring diverse topics and developing independence without compromising depth and quality is therefore essential. On the other hand, evidence shows that scientific breakthroughs often come "from the grave"[51]: only after superstars exit the field can their non-collaborators advance the field further. Our findings offer a complementary perspective, suggesting that mentorships which actively encourage independent thinking and diverse exploration can likewise facilitate the pursuit of new directions and potentially promote innovative discoveries. Taken together, this study provides guidance for early-career researchers navigating academic independence and offers evidence-based suggestions for institutions to foster innovation and support independent contributions.

# Methods

## Data source.

Our analysis is based on two distinct datasets. The first one is curated from the Academic Family Tree (AFT, Supplementary Section S1.1.1), an online website (`Academictree.org`) for collecting mentor-mentee relationships in a crowd-sourced fashion. The second data set is the OpenAlex (`https://openalex.org/`, Supplementary Section S1.1.2), a bibliographic database containing entities about authors, works (journals, conferences, etc.), affiliations, and citations. One advantage of OpenAlex over other publication databases is that all entities have been disambiguated and associated with identifiers . The name disambiguation accuracy for these authors also has been validated as reliable and unproblematic using extensive and strict procedures in prior works[46,83] (Supplementary Section S1.2.3). The AFT and OpenAlex data sets have been connected by matching the same scientists in each data set, and this matching has been validated with extensive and strict procedures[46]. The combined data of AFT and OpenAlex is taken from[46]. In this paper, we conduct our analysis on researchers in Chemistry, Physics, and Neuroscience, amounting to ~0.5 million mentor-mentee pairs, and to ~80k scientists who published ~10 million papers. We motivate our choice for the studied fields in Supplementary Section S1.2.1.

### Community detection.

The co-citation network of a pair of mentor and mentee is constructed by linking two papers if they are cited by at least one following work. The community structure of the network is detected with the fast unfolding algorithm[56], which is a heuristic method based on modularity optimization. The modularity function considered in this paper is defined as

$$Q = \frac{1}{2m} \sum_{i,j} \left[ A_{i,j} - \gamma \frac{k_i k_j}{2m} \right] \delta(c_i, c_j) \tag{3}$$

where $A_{ij}$ is an element of the adjacency matrix of the co-citation network, $k_i$ is the degree of node $i$, $m$ is the total number of links in the network, $c_i$ is the community to which node $i$ is assigned, the $\delta$ function $\delta(c_i, c_j)$ is 1 if $c_i = c_j$, and 0 otherwise. The communities are obtained when the function $Q$ is maximized. Note that $\gamma$ is a resolution parameter in $Q$. We set it as the standard modularity function, i.e., $\gamma = 1$, in this paper. We have validated the communities identified by this approach can be meaningfully associated with different research topics (Supplementary Fig. S3 and S10-S12).

### Average network distance.

In the co-citation network of papers authored by the mentor and mentee, nodes are connected if they are co-cited by at least one subsequent paper, suggesting a degree of similarity in their topical content. The longer the path between two nodes, the less similar their research interests are considered to be. Therefore, to quantify the dissimilarity in research topics between mentor and mentee, as shown in Figure 5 g-i, we first identify the shortest paths between any two nodes. We then calculate the average length ($L$) of these shortest paths for paired nodes belonging to either the mentor or the mentee.

$$L = \frac{1}{|E| \cdot |R|} \sum_{e \in E, r \in R} d(e, r) \tag{4}$$

where the $E$ and $R$ represent the sets of mentee's papers and mentor's papers in their co-citation network, respectively. $d(e, r)$ is the shortest path between node $e$ and node $r$.

Specifically, if node $e$ and node $r$ are not connected by any path, the length of the shortest path between them is considered to be equal to the length of the longest shortest path in the network, as calculated by the formula 5.

$$d(e, r) = \max_{i,j \in (E \cup R)} (d(i, j)) \tag{5}$$

### Average semantic distance.

Because community detection in co-citation networks is shaped by citation structure[48–50], using them to capture mentor-mentee topic divergence risks potential bias: when papers span a broad topical range, topic communities may reflect structural artifacts rather than genuine topical similarity. This can yield communities at inconsistent levels of topic granularity, limiting comparability across networks of different scope. To specifically mitigate possible structural bias from applying community detection on co-citation networks, we employed an independent framework that measures mentor-mentee topic proximity based solely on the titles and abstracts of their papers[84–86]. Each paper is embedded using the SPECTER2 proximity-format model[86]. Based on these embeddings, we calculated the average cosine distance $D$ between the semantic vectors of mentor's and mentee's papers to quantify their topic divergence following the procedure below.

The cosine similarity between two papers $P_i$ and $P_j$ is computed as the cosine of the angle between their semantic vectors $v_i$ and $v_j$:

$$\text{cosine similarity}(v_i, v_j) = \frac{v_i \cdot v_j}{||v_i|| \, ||v_j||} \tag{6}$$

$$d_{ij} = 1 - \text{cosine similarity}(v_i, v_j) \tag{7}$$

where $v_i \cdot v_j$ is the dot product of the vectors, and $||v_i||$ and $||v_j||$ are the Euclidean norms of the vectors.
The average cosine distance $D$ is then given by:

$$D = \frac{1}{|E| \cdot |R|} \sum_{i \in E, j \in R} d_{ij} \tag{8}$$

Note that for a given mentee-mentor pair, $E$ and $R$ denote the sets of the mentee's and mentor's papers within their co-citation network, respectively. The distance $D$ thus quantifies how distinct the semantic topics of the mentee's papers are from those of the mentor's papers, but does not rely on their publication volumes or citation network structure.

## Author contributions statement

Y.X. and Y.M. conceived the study. All authors contributed to the design of the study. Y.X. curated the datasets, developed the algorithm, performed the analysis and conducted the visualization. Y.X., Y.S., G.L. and Y.M. contributed to the interpretation of the results. Y.X., Y.S., G.L., and Y.M. wrote the manuscript. G.L. and Y.M. supervised the study.